\documentclass{LMCS}
\usepackage{amssymb,latexsym,enumerate}
\usepackage{hyperref}

\def\nek{,\ldots,}
\def\diamonds{\overleftarrow{\diamondsuit}}

\def\ovarphi{{\overline\varphi}}

\def\L{L}

\def\d1{{\diamondsuit_{1}}}
\def\bd1{\overleftarrow{\diamondsuit}_{1}}
\def\bD11{\overleftarrow{\diamondsuit}_{\leq 1}}

\def\b1{{{\square }_{1}}}
\def\bb{\overleftarrow{{\square }}}
\def\bb1{\overleftarrow{{\square }}_{1}}
\def\bB11{\overleftarrow{{\square }}_{< 1}}
\def\Models{\models_{{}_{{}_{\!\!\!\!\scriptscriptstyle
{MLO}}}}}
\def\ModelsTL{\models_{{}_{{}_{\!\!\!\!\scriptscriptstyle
{TL}}}}}

\def\L2{{\mbox{\it{Q2MLO}}}}
\def\QMLO{{\mbox{\it{QMLO}\ }}}
\def\ML1{{\mbox{$MLO^+$}}}

\def\X1l{X_1 \nek X_l}
\newcommand{\QTL}{\mathit{QTL}}
\newcommand{\TL}{\mathit{TL}}
\newcommand{\MLO}{\mathit{MLO}}
\newcommand{\FOMLO}{\mathit{FOMLO}}
\newcommand{\MITL}{\it {MITL}}
\def\power{\mbox{$\mathbb{P}$}}
\newcommand{\until}{{\bf U }}
\newcommand{\since}{{\bf S }}

\usepackage{graphicx}
{\theoremstyle{plain}

\newtheorem{theorem}{Theorem}
\newtheorem{proposition}[theorem]{Proposition}
}

\theoremstyle{definition}

\newtheorem{rem1}[theorem]{Remark}
\newtheorem{definition}[theorem]{Definition}

\def\nek{,\ldots,}
\def\diamonds{\displaystyle\diamondsuit\!\!\!\hbox{-}}

\def\ovarphi{{\overline\varphi}}

\def\Models{\models_{{}_{{}_{\!\!\!\!\scriptscriptstyle {MLO}}}}}
\def\ModelsTL{\models_{{}_{{}_{\!\!\!\!\scriptscriptstyle{TL}}}}}

\def\nek{,\ldots,}
\def\diamonds{\overleftarrow{\diamondsuit}}

\def\ovarphi{{\overline\varphi}}

\def\L{L}

\def\d1{{\diamondsuit_{1}}}
\def\bd1{\overleftarrow{\diamondsuit}_{1}}
\def\bD11{\overleftarrow{\diamondsuit}_{\leq 1}}
\def\b1{{{\square }_{1}}}
\def\bb{\overleftarrow{{\square }}}
\def\bb1{\overleftarrow{{\square }}_{1}}
\def\bB11{\overleftarrow{{\square }}_{< 1}}
\def\Models{\models_{{}_{{}_{\!\!\!\!\scriptscriptstyle{MLO}}}}}
\def\ModelsTL{\models_{{}_{{}_{\!\!\!\!\scriptscriptstyle{TL}}}}}

\def\L2{{\mbox{\it{Q2MLO}}}}
\def\QMLO{{\mbox{\it{QMLO}\ }}}
\def\ML1{{\mbox{$MLO^+$}}}

\def\X1l{X_1 \nek X_l}

\def\doi{3 (1:3) 2007}
\lmcsheading%
{\doi}
{1--11}
{}
{}
{Apr.~29, 2006}
{Feb.~23, 2007}
{}

\begin{document}
\title[Expressiveness of Metric modalities for continuous time]{Expressiveness of Metric modalities for continuous time}

\author[Y.~Hirshfeld]{Yoram Hirshfeld\rsuper{a}}   
\address{{\lsuper{a,b}}Sackler Faculty of Exact  Sciences, Tel Aviv University, Israel 69978.} 
\email{\{joram,rabinoa\}@post.tau.ac.il}  
\author[A.~Rabinovich]{Alexander Rabinovich\rsuper{b}} 

 \maketitle

\begin{abstract}
We prove a conjecture by A. Pnueli and strengthen it showing a
sequence of ``counting modalities" none of which is expressible in
the temporal logic generated by the previous modalities, over the
real line, or over the positive reals. Moreover,  there is no
finite temporal logic that can express all of them over the real
line, so that no finite metric temporal logic is expressively
complete.
\end{abstract}


\section{Introduction}
Temporal Logic based on the two modalities ``Since" and ``Until" ($\TL$) is a most popular
framework for reasoning about the evolving of a system in time. By  Kamp's theorem \cite{Kamp68} this logic  has
the same expressive power as the monadic first order predicate logic. Therefore the choice between monadic logic
and temporal logic is merely a matter of personal preference.

For discrete time these logics suffice. A properties like ``Every
$X$ will be followed promptly enough by a $Y$" can be explicitly
written once a number $k$ is chosen, and ``promptly enough" is
interpreted as: ``Within $k$ steps".

Temporal logic and the monadic logic are equivalent whether the
system evolves in discrete steps or in continuous time. But for
continuous time both logics lack the power to express properties
of the kind just described, and we must strengthen their
expressive power.

Some measure of length of time needs to be included, and the
language must be adapted to it. This is done by assuming that
there is a basic unit of length, call it ``length 1". For
predicate logic it is standard procedure to extend the language by
a name for the ``+1" function,  or for a corresponding relation.
It will then be the question which fragment of the extended
language suits our needs.

Extending temporal logic, without relating it to a corresponding
predicate logic, depended on the inventiveness and imagination of
the researchers, who created an abundance of approaches and
notions, in the work of A. Pnueli, R. Koymans, T. Henzinger and
others,\cite{Koy90,BKP85,AH92,MP93,GHR94,Wilke94,AFH96,Hen98,HFS98},
and more. Following much work, mainly by Henzinger and
collaborators, most of these approaches seem to converge to
equivalent languages.

{\sloppy  We evaluated the situation in \cite{HR99a,HR05}, when
most of the work cited was already done, and analyzed the temporal
logic in relation to predicate logic. This lead to the temporal
logic $QTL$ (Quantitative Temporal Logic), which has besides the
modalities $Until$ and $Since$ two metric modalities: $\d1(X)$ and
$\bd1(X)$. The first one says that $X$ will happen (at least once)
within the next unit of time, and the second says that $X$
happened within the last unit of time. We showed that this logic
is equivalent to the different metric temporal logics that we
found in the literature, like $MITL$, $ECL$ etc.
\cite{AH92,AFH96,Hen98,HFS98}. We will use in this paper the logic
$\QTL$ as the basic metric temporal logic, but the reader who is
acquainted with $\MITL$ or any other of the different metric
temporal logics should remember that they are equivalent to $\QTL$
in expressive power, so that the results in this work apply to
$\MITL$ and those other languages just as well.

Adding the power to say ``$X$ will be true (at least once) within
the next unit of time" is natural and necessary. There is however
no reason to believe that this gives us the required expressive
power. Is it enough, or do we need to add more modalities? If we
must add more, which ones should we choose?

A. Pnueli was the first to address these questions. He suggested
the modality $P_2(X,Y)$: $``X$ and then $Y$ will both occur in the
next unit of time". Pnueli conjectured that the modality
$P_2(X,Y)$ could not be expressed in $\MITL$ and similar logics
(we were unable to locate where this conjecture was first
published. It is attributed to Pnueli in later papers like \cite
{AH92} and \cite{Wilke94}).

$P_2(X,Y)$ was probably thought of as a natural strengthening of
the simple metric temporal logics. It can serve as a first in a
sequence of extensions of the logic, where for each natural number
$n$, we add the modality $P_n(X_1,\nek X_n)$. $P_n(X_1,\nek X_n)$
says that there is an increasing sequence of points $t_1,\nek t_n$
in the coming unit interval such that $X_i(t_i)$ holds for
$i=1,\nek n$. We call these modalities {\bf Pnueli's modalities}.

In this paper we will:
\begin{enumerate}[$\bullet$]
\item
Prove Pnueli's conjecture, that $P_2(X,Y)$ cannot be expressed in
$\QTL$ or $\MITL$.
\item
Show that none of the modalities $P_n(X_1,\nek X_n)$ can be
expressed in terms of the modalities $P_{n-1}(X_1,\nek X_{n-1})$,
so that we have a strict hierarchy of modalities.
\item
And the main result: No temporal logic with a finite set of
modalities can express all the modalities $P_n(X_1,\nek X_n)$.
\end{enumerate}

This makes clear two points: That an extension of the simple
temporal logics is necessary, and that it will not be as simple as
it was for plain temporal logic. It will require infinitely many
modalities, and the proper choice is an intriguing question. We
hope to address it soon. Note that in predicate logic the
expressive power of formulas grows with the increasing of their
quantifier depth. In temporal logic there are no quantifiers, and
formulas become more complex due to increase in the nesting depth
of the modalities that they mention. Kamp showed that for the
simplest logic of order iterating the modal operations can replace
the complex use of quantifiers. Our result, together with previous
evidence (see \cite{Rab02}) suggests that this was a lucky
peculiarity of the first-order monadic logic of linear order, and
that it cannot be expected to hold for stronger logics.

The main result, that no finite temporal logic can be complete, is
not an exact formal claim, until we specify which source of
modalities we have in mind. We will specify a natural extensive
monadic logic of order, that includes the ``+1" function in its
vocabulary. The formal claim will concern the modalities which are
definable in this logic. When stated formally the result seems
even stronger, as it states not just the incompleteness of
temporal logics with finitely many modalities, but also of logics
with infinitely many modalities, which are defined using bounded
quantifier depth.

To state the formal result we define the {\bf counting modalities}
$C_n(X)$ which are a simple instance of Pnueli's modalities.
$C_n(X)$ says that {\em $X$ will hold at least at $n$ points
within the next unit of time}. $C_n(X)$ is a simple instance of
the Pnueli modality $P_n(X,\nek X)$. Our main theorem is the
following:
\begin{enumerate}[$\bullet$]
\item

Let $L$ be {\em second order} monadic logic of order, together
with the predicate $B(t,s)$ which says that $s=t+1$. The
modalities $C_n(X)$ are expressible in this logic, but no temporal
logic with a finite {\em or infinite} family of modalities which
are defined by formulas with bounded quantifier depth can express
all the modalities $C_n(X)$,  {\em over the full real line $R$}.
\end{enumerate}

\begin{rem1}
Note that our proof applies only to the entire real line. We
conjecture that the same claim holds for the positive real line
$R^+$, but the attempts to prove it became too cumbersome to carry
on.
\end{rem1}
 It is well known that in the theory of order, to
express the fact that there is a large number of points with a
given property, requires formulas with large quantifier depth. We
emphasize that the main theorem is not of this nature. Thus in
pure temporal logic the two modalities ``Until" and ``Since"
suffice to express for every $n$ the fact that there are $n$
points in the future that satisfy $X$. Less trivial but true, is
the fact that in $\QTL$ with its four modalities, for every $n$
there is a formula that says about a point $t$ that $X$ will be
true for the length of the interval $(t+n,t+n+1)$ \cite{HR04}. The
nesting of the modalities does for the temporal formulas what the
quantifier depth does for the predicate logic formulas. The
theorem says that in the general case modality nesting is strictly
weaker than quantifier depth, and that no temporal logic will be
expressive enough unless it has infinitely many modalities,
defined using definitions of unbounded complexity, in terms of
quantifier depth.

 The paper is divided as follows: In section 2 we recall the definitions and the
 previous results concerning
the continuous time logics. In section 3 we prove Pnueli's
conjecture and its generalization, that the modalities $C_i$
create a strictly increasing family of logics. In section 4 we
discuss the more general and abstract result:
 that no temporal
logic based on modalities with finite quantifier depth can express
all the modalities $C_n$.

\section{Monadic Logic and Quantitative Temporal Logic}
\subsection{MLO - Monadic Logic of Order}

The natural way to discuss systems that evolve in time is
classical predicate  logic. The language has a name for the order
relation of the time line, and a supply of unary predicate names
to denote a properties that the system may or may not have at any
point in time. Hence:

{\bf The syntax of the monadic predicate logic of order - MLO\/}
has in its vocabulary {\it individual\/} (first order) variables
$t_0,t_1\nek$ monadic {\it predicate \/} variables  $X_0,X_1\nek$
 and one binary
relation $<$ (the order). {\bf Atomic formulas\/} are of the form
$X(t)$, $ t_1=t_2$ and $ t_1<t_2$.  {\bf Well formed formulas\/}
of the monadic logic $\MLO$ are obtained from atomic formulas
using Boolean connectives $\neg, \vee, \wedge,\to$ and the (first
order) quantifiers $\exists t$  and $\forall t$  and the
(second-order) quantifiers $\exists X$  and $\forall X$. The
formulas which do not use $\exists X$  and $\forall X$ are called
first-order $\MLO$ formulas ($\FOMLO$). Note that $\FOMLO$
formulas may contain free monadic predicate variables, and they
will be assigned to particular predicates in a structure.

{\bf A structure for $\MLO$\/} is a tuple $M=\langle
A,<,P_1,\dots,P_n\rangle $, where $A$ is a set linearly ordered by
the relation $<$, and  $P_1,\cdots,P_n$, are one-place predicates
(sets) that correspond to the predicate names in the logic. We
shall use the simple notation $\langle A,<\rangle$ when the
particular predicates are not essential to the discussion.

 The main models are: the {\bf continuous canonical model}
$\langle R^+,<\rangle$, the non-negative real line, and  the {\bf
discrete canonical model} $\langle N, <\rangle$, the naturals.

As is common we will use the assigned formal names to refer to
objects in the meta discussion. Thus we will write:
 $$M\models\varphi[t_1\nek t_k; X_1\nek X_m]$$
 where $M$ is a structure, $\varphi$ a formula, $t_1,\cdots, t_k$
 elements of $M$ and $X_1\nek X_m$ predicates in $M$, instead of
 the correct but tedious form:
 $$ M,\tau_1,\ldots ,\tau_k ; P_1,\ldots,P_m
\Models\varphi(t_1,\ldots, t_k ;X_1\nek X_m),$$ where
$\tau_1,\ldots ,\tau_k$ and $P_1\cdots, P_m$ are names in the
metalanguage for elements and predicates in $M$.

\subsection{Temporal Logics} \label{sect:tl}
Temporal logics evolved in philosophical logic and were enthusiastically embraced by a large body of computer
scientists. It uses logical constructs called ``modalities" to create a language that is free from variables and
quantifiers. Here is the general logical framework to define temporal logics:

{\bf The syntax of the Temporal Logic $TL(O^{(k_1)}_1,\dots,
O_n^{(k_n)} ,\dots)$\/} has in its vocabulary {\it monadic
predicate names\/} $P_1,P_2,\ldots$ and a sequence of {\it
modality names\/} with prescribed arity, $O^{(k_1)}_1,\dots,
O_n^{(k_n)} ,\dots $ (the arity notation is usually omitted). The
formulas of this temporal logic are given by the grammar:
\[\varphi~::=True|~P~|~\neg\varphi~|~\varphi\wedge\varphi~|~ O^{(k)}(\varphi_1,\cdots,\varphi_k)\]
A temporal logic with a finite set of modalities is called a
finite (base) temporal logic.

{\bf Structures for TL} are again linear orders equipped with monadic
predicates $M=\langle A,<, P_1,P_2, \dots, P_n\rangle$, where the
predicate $P_i$ are those which are mentioned in the formulas of
the logic. Every modality $O^{(k)}$ is interpreted in every
structure $M$ as an operator $O^{(k)}_M:[\power(A)]^k\to
\power(A)$ which assigns ``the set of points where
$O^{(k)}[S_1\nek S_k]$ holds" to the $k$-tuple $\langle S_1\nek
S_k\rangle\in \power(A)^k$. (Here $\power$ is the power set
notation, and $\power({A})$ denotes the set of all subsets
 of  $A$.) Once every
modality corresponds to an operator the semantics is defined by
structural induction:

\begin{enumerate}[$\bullet$]
\item
for atomic formulas: $\langle M,t \rangle \ModelsTL P\quad {\rm
iff}\quad t\in P$.
\item
for Boolean combinations the definition is the usual one.
\item
for  $O^{(k)}(\varphi_1,\cdots,\varphi_k)$
$$\langle M,t\rangle \ModelsTL O^{(k)}(\varphi_1,\cdots,\varphi_k)\quad {\rm
iff}\quad t\in O^{(k)}_M(A_{\varphi_1},\cdots,A_{\varphi_k})$$
where $A_{\varphi}~=~\{~\tau~:~\langle M,\tau \rangle \ModelsTL
\varphi~\}$ (we suppressed predicate parameters that may occur in
the formulas).
\end{enumerate}
We are interested in a more restricted case; for the modality to
be of interest the operator $O^{(k)}$ should reflect some intended
connection between the sets $A_{\varphi_i}$ of points satisfying
$\varphi_i$ and the set of points $O[A_{\varphi_1},\ldots,
A_{\varphi_k}]$. The intended meaning is usually given by a
formula in an appropriate predicate logic:

{\bf Truth Tables}: A formula $\overline{O}(t_0,  X_1,\ldots X_k)$
in the predicate logic $L$ is a {\it Truth Table} for the modality
$O^{(k)}$ if for every structure $M$

$$O_M(A_1\nek A_k)=\{\tau~:~
M\Models\overline{O}[\tau, A_1\nek A_k]\}\ .$$
 The modalities {\em until} and {\em since} are most commonly  used
in temporal logic for computer science. They are defined through the following truth tables:

\begin{enumerate}[$\bullet$]

\item
The modality $X\until\ Y$, ``$X$ {\it until \/} $Y$", is defined
by
$$\psi (t_0,X,Y)\equiv \exists t_1(t_0<t_1\wedge
Y(t_1)\wedge\forall t(t_0<t<t_1\to X(t))).$$
\item
The modality $X\since\ Y$, ``$X$ {\it since \/} $Y$", is defined
by $$\psi (t_0,X,Y)\equiv \exists t_1(t_0>t_1\wedge
Y(t_1)\wedge\forall t(t_1<t<t_0\to X(t))).$$
\end{enumerate}
If the modalities of a temporal logic have truth tables in a
predicate logic then the temporal logic is equivalent to a
fragment of the predicate logic. Formally:
\begin{proposition}\label{prop-tl-to-fo}
If every modality in the temporal logic  $\TL$ has a truth table
in the logic $\MLO$ then to every formula $\varphi(X_1,\ldots,
X_n)$ of $\TL$ there corresponds effectively (and naturally) a
formula $\ovarphi(t_0,X_1,\ldots X_n)$ of $\MLO$ such that for
every $M$, $\tau\in M$  and predicates $P_1,\ldots, P_n$
$$\langle M,\tau, P_1,\ldots, P_n\rangle \ModelsTL\varphi
\quad {\rm iff  }\quad \langle  M,\tau,P_1,\ldots ,P_n\rangle
\Models\ovarphi\ .$$
\end{proposition}
In particular the temporal logic $TL(\until, \since )$ with the
modalities ``until" and ``since" corresponds to a fragment of
first-order $\MLO$ ($\FOMLO$).

The two modalities $\until$ and $\since$ are also enough to
express all the formulas of first-order $MLO$ with one free
variable:
\begin{theorem}
 \emph{(\cite{Kamp68,GPSS80})} The temporal
logic $TL(\until,\since)$ is expressively complete for $\FOMLO$
over the two canonical structures: For every formula of $\FOMLO$
with at most one free variable, there is a formula of  $TL(\until
, \since )$, such that the two formulas are equivalent to each
other, over the positive integers (discrete time) and over the
positive real line (continuous time).
\end{theorem}

\subsection{QTL - Quantitative Temporal Logic}

The logics $\MLO$ and $TL(\until,\since)$ are not suitable to deal
with quantitative statements like ``$X$ will occur within one unit
of time". In  \cite{HR99,HR99a,HR04} we introduced the {\em
Quantitative Temporal Logic}, adding to $\TL$ the modalities
$\diamondsuit_1X$ ($X$ will happen within the next unit of time)
and $\diamonds_1X$ ($X$ happened within the last unit of time):

\begin{definition}[Quantitative Temporal Logic]
$QTL$, {\em quantitative temporal logic} is the logic $TL(\until , \since )$
enhanced by the two modalities:  $\diamondsuit_1X$  and $\diamonds_1X$. These modalities are defined by the
tables with free variable $t_0$:
$$\diamondsuit_1X:\qquad\exists t((t_0<t<t_0+1)\wedge
X(t))\leqno(3)$$
$$\diamonds_1X:\qquad\exists t(( t<t_0<t+1)\wedge
X(t))\ .\leqno(4)$$
\end{definition}
$\QTL$ was the latest in a list of metric logics for continuous
time, developed over approximately 15 years. When interpreted
carefully all these logics are equivalent. We refer the reader to
\cite{Koy90,BKP85,AH92,MP93,GHR94,Wilke94,AFH96,Hen98,HFS98} for
some of the previous work.
\vfill\eject
The novelty in our approach was the close connection with metric
monadic logic, the replacement of all the automata theory
arguments by plain logic and model theory arguments. Most
significant however was the fact that our treatment and our
results applied uniformly to the class of systems with finite
variability and to the class of all systems. In contrast, in the
previous work, and in particular in the papers cited above,
systems without finite variability could not be defined as the
semantics for the logic. Naturally the decidability and complexity
results did not apply to systems without finite variability (nor
could the automata approach be adapted to the general case once
the definition includes general systems).

We proved in \cite{HR99a} and \cite{HR05} that:
\begin{enumerate}[(1)]
\item

$\QTL$ consumes the different decidable metric temporal logics
that we found in the literature, including $MITL$, $ECL$ etc.

\item
There is a natural fragment $\QMLO$ (quantitative monadic logic of
order), of the classical monadic logic of order with the $+1$
function, that is equal in  expressive power to $\QTL$.
\item
The {\em validity and satisfiability problem for this logic is
decidable}, whether we are interested in systems with {\em finite
variability}, or in all systems evolving in time (a system has
finite variability if it changes only at finitely many points, in
any finite interval of time).
\end{enumerate}
For the special case of systems with finite variability these
results (but not the proof methods) are in
\cite{Koy90,BKP85,AH92,MP93,GHR94,Wilke94,AFH96,Hen98,HFS98},
regarding different logics (future, or full), and different
semantics (point sequence and interval sequence).

\section{ Modalities which are not expressible in $\QTL$}

The simple metric temporal logic $\QTL$ looks very natural. The
main question is if it is as expressive as is needed. A. Pnueli
suggested a natural modality, and conjectured that it could not be
expressed in the simple metric temporal logics of the previous
section. This was the modality that we denote by $P_2(X,Y)$, which
says that $X$ and then $Y$ will be true at two points in the next
unit of time. If the conjecture is confirmed then we have a
natural modality to add to the logic. Moreover, there are
$P_3(X,Y,Z)$ and $P_n(X_1,\nek X_n)$ waiting to be considered as
an addition, if they are not redundant.

We attend these questions and we will prove first Pnueli's
conjecture, that $P_2(X,Y)$ is not expressible in $\QTL$, and then
that there is a proper hierarchy of Pnueli modalities that can be
added to strengthen the logic.

\begin{definition}\hfill
\begin{enumerate}[(1)]
\item
The {\em counting modalities} are the modalities $C_n(X)$ for
every $n$ which state that $X$ will be true at least at $n$ points
within the next unit of time.
\item
The {\em Pnueli modalities} are the modalities $P_n(X_1\nek X_n)$
for every $n$ which state that  there is an increasing sequence of
points  $t_1,\nek t_n$ in the unit interval ahead, such that for
$i=1,\nek n$, $t_i$ is in $X_i$.
\end{enumerate}
\end{definition}

Pnueli's conjecture is proved by the following theorem:
\begin{theorem} \label{und} The modality $C_2 (X)$ is not expressible
in $\QTL$.
\end{theorem}
\begin{proof}
Let $M$ be the real non negative line with the predicate $P(t)$
that is true exactly at the points $ n\cdot \frac{2}{3}$ for all
natural numbers n. Let us call the following four predicates:
$P,\neg P,True,False$ the {\bf trivial predicates}. We show by
structural induction that for every statement $\varphi$ of $\QTL$
there is a point $t_{\varphi}$ such that from this point on
$\varphi$ is equivalent to one of the trivial predicates.
\begin{enumerate}[$\bullet$]
\item
this is trivially true for atomic statements.
\item
 The collection of truth sets for the four trivial predicates
is closed under Boolean combinations. Therefore the set of formulas satisfying our claim is closed under the
Boolean connectors.
 \item
Assume now that $\varphi=(\theta~ \until~\psi$) and $t_0$ is a
point beyond which both $\theta$ and $\psi$ are equivalent to one
of the trivial predicates. We check the different possibilities
for the truth value of $\varphi$ at a point $t$ beyond $t_0$. If
$\theta$ is equivalent to $P$ or to $False$ then $\varphi$ is
false. If $\theta$ is equivalent to $\neg P$ or to $True$ then
$\varphi$ is true if $\psi$ is equivalent to either of $P$ ,$\neg
P$ or $True$, and $\varphi$ is false if $\psi $ is equivalent to
$False$. In every case $\varphi$ is equivalent either to $True$ or
to $False$.
\item
 For $\varphi=(\theta~ Since~\psi$) we need only a
minor modification: Let $t_1$ be an even integer beyond $t_0$ (so
that $P$ is true at $t_1$). Then for points beyond $t_1$ $\varphi$
is true if $\theta\equiv True$ and $\psi$ occurred at $t_1$ or
earlier,  or if $\theta\equiv \neg P$ and $\psi$ is equivalent to
any of the special predicates except $False$ (the choice of $t_1$
ensures the case that $\psi\equiv P$) in all other cases
$\varphi\equiv False$.
\item
Assume that $\varphi=\d1\theta$ and from $t_0$ on $\theta$ is equivalent to one of the four trivial predicates.
If $\theta$ is equivalent to $False$ then $\varphi$ is equivalent to $False$ from $t_0$ on. In the other three
cases  $\varphi$ is equivalent to $True$ from $t_0$ on.
\item
A similar argument works when $\varphi=\bd1\theta$.
\end{enumerate}
On the other hand the statement $C_2(P)$ is false at any point in
the interval $(n,n+1/3)$ if $n$ is even and it is true at any
point in the interval $(n,n+1/3)$ if $n$ is odd. This shows that
$C_2(P)$ is not equivalent to any $\QTL$ formula.
\end{proof}

The method of the proof can be adapted to show that the Pnueli
modalities yield a strictly monotone sequence of temporal logics:

\begin{theorem}\label{hierarchy} The modality $C_{n} (X)$ is not expressible
in the logic $QTL(P_2,\cdots,P_{n-1})$.
\end{theorem}

\begin{proof}
Let $M$ be the real non negative line with the predicate $P(t)$
that is true exactly at the points $ k\cdot \frac{2}{2n-1}$ for
all natural numbers k.  Call again the following four predicates:
$P,\neg P,True,False$ the {\bf trivial predicates}, and as before
show that every formula of $QTL(P_2\cdots,P_{n-1})$ is equivalent
from some point on to a trivial predicate. The proof remains the
same except for the additional induction step, where we assume
that the claim is true for $\varphi_1\nek \varphi_{n-1}$. and we
must show that it holds for $\psi=P_{n-1}(\varphi_1\nek
\varphi_{n-1})$. By assumption there is some point from which on
$\varphi_1\nek \varphi_{n-1}$ are trivial. If any of them is
$False$ then $\psi$ is false from there on. Otherwise $\psi$ is
$True$ from there on, because at any point there are $n-1$ points
of $P$ in the future unit, and between any two, there are
infinitely many points that satisfy $\neg P$ or $True$.

On the other hand $C_{n} (P)$ is always true on the interval
$(k,k+\frac{1}{2n-1})$ if $k$ is even, and false on the interval
if $k$ is odd.
\end{proof}

\begin{rem1} We proved the two theorems for the positive real line.
I.e, for continuous time with a first moment. The same proof
applies to the full real line, to the set of rational numbers or
to the set of positive rational numbers.
\end{rem1}

\section{The incompleteness of temporal logic with finitely many modalities}

The hierarchy
$$\TL<\QTL<\QTL(P_2)<\cdots<QTL(P_2,\cdots,P_n)<\cdots$$ raises the
suspicion that it will be difficult to find a finite temporal
logic that includes all these logics. In this section we will
prove that  it is indeed impossible. To be precise:

\begin{theorem}\label{main}
Let $L$ be the {\em second order monadic logic} of order, with an
extra predicate $B(t,s)$ that is interpreted on the {\em whole
real line} as $s=t+1$. Let $L_1$ be a temporal logic with possibly
infinitely many modalities, for which there is a natural number
$m$ such that all the modalities have truth tables in $L$, with
quantifier depth not larger than $m$. Then there is some $n$ such
that $C_n(X)$ is not equivalent {\em over the real line} to any
$L_1$ formula.

\end{theorem}

\bigskip

Before we start to work toward the proof we make the following
remarks.
\begin{enumerate}[(1)]
\item
Second order monadic logic of order with the $+1$ function is a
much stronger logic than is usually considered when temporal
logics are defined. All the temporal logics that we saw in the
literature are defined in a fragment of  monadic logic, with a
very restricted use of the $+1$ function. All the decidable
temporal logics in the literature remain decidable when we add the
counting modalities $C_n(X)$ \cite{HR04}. On the other hand second
order monadic logic is undecidable over the reals even without the
$+1$ function \cite{She75}. When the $+1$ function is added even a
very restricted fragment of {\em first order} monadic logic of
order is undecidable over the positive reals.

\item
The theorem says that there is no finite set of modalities {\em
defined in this language} that generates all the counting
modalities (and possibly more). It does not exclude the
possibility that a finite set of modalities which are not defined
in this logic is (at least) as strong as $\QTL(P_2\nek
P_n,\ldots)$. We state it as an open problem:

\smallskip
\noindent{\bf Question:} Is there a finite temporal logic that
includes all the modalities $P_n(X_1\nek X_n)$, if we do not require
that the modalities are are defined by truth tables?

\item
On the other hand the theorem does not just say that $\QTL(P_2\nek
P_n,\ldots)$ is not a sublogic of a logic with finitely many
modalities.  It is not even a sublogic of an {\em infinite
temporal logic} whose modalities are defined with bounded
quantifier depth, in a strong predicate logic.

\end{enumerate}
The proof of theorem \ref{main} involves some more notations, and
some steps that are accumulated in two more theorems.

 We will assume that $P$ is the only non variable unary predicate name in $L$, and we
concentrate on a class of simple models in the language: For each
integer $k>0$ let $M_{k}$  be the full real line $R$ with $P(t)$
occurring at the points $m\frac{1}{k}$  for every integer $m$
(positive, negative or zero).

\subsection*{Terminology}
\begin{enumerate}[(1)]
\item From now on whenever we say ``a model" we mean $M_{k}$ for some $k$.

\item The four formulas $\{True, False, P, \neg P\}$ will be
called the {\em trivial} temporal logic formulas, and the formulas
$\{True, False, P(t_0), \neg P(t_0)\}$ will be called the {\em
trivial} first-order  formulas.

\item
We say that {\em $t$ is a $P$-point} or that {\em $t$ is in  $P$}
if $P(t)$ is true.

\end{enumerate}

The choice of the models limits the expressive power of monadic
logic:

\begin{theorem}\label{triviality}
Every formula of  second order monadic logic of order, with the
unary predicate $P$ and with the extra predicate $B(t,s)$ with one
free element variable $t_0$ and no  free predicate variable is
equivalent in every model to one of the four trivial formulas.

\end{theorem}

\proof 
Let $M_k$ be given. We show that for every two points $t<s$ in
$P$, there is an automorphism of the model that maps $t$ to $s$
and for every two points not in $P$ there is such an automorphism.
This will prove that every formula obtains the same truth value on
all points in $P$ and the same truth value on all points not in
$P$. Therefore the formula is true either everywhere, or only on
points in $P$, or only outside of $P$, or nowhere.

 \begin{enumerate}[(1)]
  \item
 the mapping $G(t)=t+r$ is an automorphism if $r$ is a multiple of $\frac{1}{k}$. This shows in particular that
 every formula obtains a fixed truth value on all points in $P$.
 \item
For every $0<t<s<\frac{1}{k}$ there is a monotone bijection
$h(v):(0,\frac{1}{k})\rightarrow (0,\frac{1}{k})$ such that
$h(t)=s$. Every real $t$ can be written in a unique way as
$t=\frac{m}{k}+\tau$ where $m$ is integer and
$0\leq\tau<\frac{1}{k}$, and the bijection extends to an
automorphism of the model defining
$H(\frac{m}{k}+\tau)=h(\tau)+\frac{m}{k}$. This shows that every
formula obtains the same truth value on the interval
$(0,\frac{1}{k})$.
\item
Finally if $t=\frac{m}{k}+\tau$ where $m$ is integer and
$0\leq\tau<\frac{1}{k}$ then every formula has the same truth
value on $t$ and on $\tau$, so that it is fixed on the complement
of $P$.\qed
\end{enumerate}
%

Thus every formula is equivalent to one of the four trivial
formulas in every model. Note that it is not necessarily the same
trivial formula in the different models. For example $ C_{k}(P)$
is equivalent to $\neg P$ in $M_{k}$ and to $True$ in $M_{k+1}$.
We aim to show that in a temporal logic based on modalities
defined by formulas of bounded quantifier depth, there are always
pairs of models for which any formula is equivalent in both to the
same trivial formula.

We denote by $L_n$  the set of formulas of   quantifier depthes no
more than $n$ of the second order monadic logic of order  with the
extra predicate $B(t,s)$.
%
%
%
 We
denote by $TL_n$ the temporal logic with all the modalities that
have a truth table in $L_n$. Every formula of $TL_n$ is equivalent
in every model to one of the trivial formulas.

Let now $\varphi(t_0, X_1,\dots, X_k)$ be a formula of $L_n$. Its
{\em localizations} are the $4^k$ formulas
$\varphi(t_0,T_1,\dots,T_k )$ , where $T_1,\dots,T_k$ vary over
all possible combinations of trivial predicates $P,\neg P, True,
False$. Note that the localizations have the same quantifier depth
as the original formula, and that beside the predicate name $P$
there are no predicate variables in the localizations.

Here is the main theorem. Item (4) is the promised result:

\begin{theorem}\label{technical}

Let $n$ be given and let $S_n$ be the collection of the
localizations of all the (truth tables of) modalities in $TL_n$.
Then:
\begin{enumerate}[\em(1)]
\item
Let $M=M_k$ and $M'=M_l$ be models. If there is a formula of
$TL_n$ which is equivalent to different trivial formulas in $M$
and in $M'$ then there is also such a formula in $S_n$.
\item
There are finitely many formulas in $S_n$ such that every formula
of $S_n$ is logically equivalent to one of them.
\item
There are models $M_k$ and $M_l$   for which any formula of $TL_n$
is equivalent to the same trivial formula in both models.
\item
There are integers $k$ such that $C_k(X)$ is not expressible in
$TL_n$.
\end{enumerate}
\end{theorem}
\proof\hfill 
\begin{enumerate}[(1)]
\item
We assume that every formula in $S_n$ is equivalent to the same
trivial formula in $M$ and in $M'$ and we prove by structural
induction that the same is true for every formula $\varphi$ of
$TL_n$. If  $\varphi$ s atomic it is $P$ or $True$. If it is a
Boolean combination of simpler formulas then the property is
trivially inherited from the simpler formulas. It remains to check
the case where $\varphi= O^{(k)}(\varphi_1,\cdots,\varphi_k)$. By
assumption for $i=1\cdots k$, $\varphi_i$ is equivalent in both
models to the same trivial formula $T_i$. If now
$\overline{O}(t_0, X_1,\ldots X_k)$ is the truth table for the
modality $O^{(k)}(X_1,\cdots,X_k)$  then $\varphi$ is equivalent
in both models to $\overline{O}(t_0,  T_1,\ldots T_k)$. This is a
formula in $S_n$ and therefore expresses the same trivial
predicate in both models.
\item
This is a well known simple property of logics with  finite
relational  signature: For every  $n$ and $m$ there are finitely
many {\em quantifier free} formulas such that every quantifier
free formula with variables among $v_1,\cdots,v_n$ and
$Y_1,\cdots,Y_m$, is logically equivalent to one of them.
Consequently for every $q,~m$ and $n$ there is a finite number of
formulas such that every formula of quantifier depth $n$ (of both
first and second order variables) and in the free variables
$v_1,\cdots,v_q$ and $Y_1,\cdots,Y_m$, is logically equivalent to
one of them. $S_n$ is a special case where $q=1$ and $m=0$.
\item
Let $\phi_1,\dots,\phi_r$ be a list of formulas such that every
formula of $S_n$ is logically equivalent to one of the formulas in
the list. We partition the class of models into $4^r$ classes
according to which trivial predicate is defined by these $r$
formulas. Then at least one class is infinite. Every formula in
$S_n$ is equivalent in all the models in this class to the same
trivial formula. By (1) any formula of $TL_n$ is equivalent to the
same trivial predicate in all models in this class.
\item
Let now $M_k$ and $M_l$ be in this infinite class with $k<l$.
$C_k(P)$ is equivalent to $\neg P$ in $M_k$ and to $True$ in
$M_l$.  However, by (3) and our choice of $k$ and $l$, every
formula of $TL_n$ is equivalent to the same trivial formula in
both  models $M_k$ and $M_l$.  This shows that $C_k$ is not
equivalent to any $TL_n$ formula.\qed
\end{enumerate}

\medskip
\noindent {\bf The proof of theorem \ref{main}} is just item 4 in theorem
\ref{technical}.
\medskip

Some remarks are in order:
\begin{enumerate}[(1)]
\item
The theorem speaks about the whole real line, and not about its
non negative part $R^+$. We believe that it is true also for the
model of non negative reals, but we did not pursue the proof,
which is complicated by everything that can be said about $0$, and
therefore about every particular $n$, and about every particular
interval $(n,n+1)$. We state it as a question:

\smallskip\noindent{\bf Question:} Is the theorem above true when the real line $R$
is replaced by its non negative part $R^+$?

\item
The completeness of the real line was never used and the same
proof works for the model of the rational numbers.
\item
Adding just a ``+1" function is weaker than adding the functions
``+q" for every rational number $q$. Let us denote by $L_Q$ the
monadic logic of order with predicates $B_q(t,s)$ for every
rational umber $q$, to express the relation $s=t+q$. The proof of
the theorem will not apply if we replace $L$ by $L_Q$, and even
modalities with truth tables of quantifier depth 2 distinguish
between any two models $M_k$ and $M_r$ in our class. On the other
hand just as before no {\em finite} temporal logic defined in this
logic can express all the counting modalities. This is the case
because any finite number of modalities defined in $L_Q$ involves
only finitely many rational numbers in the formulas $B_q(t,s)$
that occur in the definition. Let $q_0$ be a rational such that
all of these rational are whole integers of $q_0$. Then we can
repeat the proof above with $q_0$ replacing $1$.  We do not know
whether infinitely many modalities defined with bounded quantifier
depth may suffice:

\medskip
\noindent {\bf Question:} Is the theorem above true when the
predicate logic $L$ is replaced by $L_Q$?

\medskip

\end{enumerate}

\section{Conclusion}
Temporal logic does not have quantifiers, and formulas become more
complex only due to deeper nesting of modalities. By Kamp's
theorem this suffices to capture the expressive power that is
achieved in predicate logic by quantifier depth, and the pure
temporal logic with its two modalities is as expressive as the
pure monadic logic of order. We proved that in the more general
setting of a metric temporal logic, nesting of modalities is
strictly weaker than nesting of quantifiers. Only a set of
modalities defined with unbounded quantifier depth can capture all
the counting modalities  $C_n(X)$.

\vskip-15 pt


\begin{thebibliography}{10}
\bibitem{AFH96} R. Alur, T. Feder, T.A.
Henzinger.  The Benefits of Relaxing Punctuality. Journal of the
ACM 43  116-146, (1996).

\bibitem{AH92} R. Alur, T.A. Henzinger.
Logics and Models of Real Time: a survey.  In Real Time:  Theory
and Practice.  Editors de Bakker et al. LNCS 600
 74-106, (1992).

\bibitem{BKP85} H. Barringer, R. Kuiper, A. Pnueli. A really abstract concurrent
model and its temporal logic. Proceedings of the 13th annual
symposium on principles of programing languages
 173-183, (1986).

\bibitem{EF91} H.D. Ebbinghaus, J. Flum,
Finite Model Theory. Perspectives in mathematical logic, Springer
(1991).

\bibitem{GHR94} D.M. Gabbay, I. Hodkinson,
M. Reynolds.  Temporal Logics volume 1.  Clarendon Press, Oxford
(1994).

\bibitem{GPSS80} D.M. Gabbay, A. Pnueli,
S. Shelah, J. Stavi.  On the Temporal Analysis of Fairness.  7th
ACM Symposium on Principles of Programming Languages.  Las Vegas
163-173, (1980).
\bibitem{Hen98} T.A. Henzinger.
It's about time: real-time logics reviewed. In Concur 98, Lecture
Notes in Computer Science 1466, pp. 439-454, (1998).

\bibitem{HFS98} T.H Henziger, J.F Raskin, P.Y Schobbens. The
regular real time languages. ICALP98, pp. 580-591,  (1998).

\bibitem{HR99}
Y. Hirshfeld and A. Rabinovich, A Framework for Decidable Metrical
Logics.
 In Proc.  26th ICALP Colloquium, LNCS vol.1644, pp. 422-432, Springer Verlag, (1999).
\bibitem{HR99a}Y. Hirshfeld and A. Rabinovich.
Quantitative Temporal Logic. In Computer Science Logic 1999, LNCS
vol. 1683, pp. 172-187, Springer Verlag (1999).
\bibitem{HR04} Y. Hirshfeld and A. Rabinovich, Logics for Real Time: Decidability and Complexity. Fundam. Inform.
62(1):1-28 (2004).
\bibitem{HR05} Y. Hirshfeld and A. Rabinovich, Timer formulas and decidable metric
temporal logic. Information and Computation Vol 198(2), pp.
148-178, (2005).
\bibitem{Kamp68} H. Kamp. Tense Logic
and the Theory of Linear Order.  Ph.D. thesis, University of
California L.A. (1968).
\bibitem{MP93} Z. Manna, A. Pnueli. Models for reactivity. Acta informatica
30:609-678, (1993).
\bibitem{Koy90}
R. Koymans. Specifying Real-Time Properties with Metric Temporal
Logic. Real-Time Systems 2(4):255-299, (1990).

\bibitem{Rab02} A.  Rabinovich. Expressive Power of Temporal Logics
In Proc. 13th Int. Conf. on Concurrency Theory, vol. 2421 of
Lecture Notes in Computer Science, pp. 57--75. Springer, (2002).





\bibitem{She75}
S. Shelah.
\newblock The monadic theory of order.
\newblock  {\em Ann. of Math.}, { 102}, pp. 349-419, (1975).



\bibitem{Wilke94} T. Wilke.  Specifying
Time State Sequences in Powerful Decidable Logics and Time
Automata.  In Formal Techniques in Real Time and Fault Tolerance
Systems. LNCS 863, pp.  694-715, (1994).

\end{thebibliography}
\end{document}